\documentstyle[epsf,preprint,aps]{revtex}
\draft
\preprint{SNUTP 00/029}
\def\cov{\bigtriangledown}
\begin{document}
\title{\Large\bf Gravity in the Einstein-Gauss-Bonnet Theory 
with the Randall-Sundrum Background}
\author{Jihn E. Kim and Hyun Min Lee} 
\address{ Department of Physics and Center for Theoretical Physics, 
Seoul National University,
Seoul 151-742, Korea}
\maketitle

\begin{abstract} 
We obtain the full 5D graviton propagator in the Randall-Sundrum 
model with the Gauss-Bonnet interaction. From the 
decomposition of the graviton propagator on the brane, 
we show that localization of gravity arises in the presence
of the Gauss-Bonnet term. We also obtain the metric perturbation for 
observers on the brane with considering the brane bending and compute
the amplitude of one massless graviton exchange. For the positive definite 
amplitude or no ghost states, the sign of the Gauss-Bonnet coefficient 
should be negative in our convention, which is 
compatible with string amplitude computations.
In that case, the ghost-free condition is sufficient for obtaining
the Newtonian gravity.  For a vanishing Gauss-Bonnet coefficient, 
the brane bending allows us to reproduce the correct graviton 
polarizations for the effective 4D Einstein gravity.\\
\end{abstract}



\newpage
\section{introduction}

The Randall-Sundrum(RS)\cite{RS1,RS2} model has drawn much 
attention recently because of several fascinating features
related to the new attempt toward the hierarchy solution. 
The RS I model\cite{RS1} is composed of two branes of the 
opposite tensions located at the orbifold fixed points in 
5D non-factorizable geometry (or a slice of $AdS_5$), which 
was proposed to explain the gauge hierarchy between the Planck 
and the electroweak scales.  Here, the negative tension 
brane is considered as the visible brane. On the other hand, 
the RS II model\cite{RS2} 
has a single brane with a positive tension in the 5D $\it {warped}$ 
geometry (or a glued patch of two $AdS_5$ spaces with a UV cutoff), 
where it is possible to get the localization of gravity on the 
brane even for non-compact extra dimension.

However, the RS I model is cosmologically problematic for several 
reasons. Firstly, the Hubble parameter, 
$H\propto \rho^{1/2}$, in the early 
universe and the negative tension brane as the visible brane
may be in conflict, since our universe would give 
rise to an imaginary Hubble parameter in a later cosmology at low 
temperatures\cite{hubble}. Second, there is no mechanism of 
radius stabilization in the RS I model itself\cite{cosmo}. 
Third, it is not obvious whether the universe with matter tends 
to the RS limit necessarily. For solving some of these problems, 
there was a proposal of radius stabilization with a bulk scalar 
coupled to branes\cite{wise,radcos}, and introduction of
the Gauss-Bonnet term\cite{KKL}. Also, there exists a proposal 
interpreting the positive tension brane as the visible brane
and the negative tension brane as the intermediate scale brane~\cite{mub}
to obtain the $\mu$-term in supergravity through nonrenormalizable
superpotential~\cite{mu}. In the last case, supersymmetry is necessary, 
which is redundant for a solution of the gauge hierarchy problem but
maybe inevitable if some problems such as the unification of
of the particle forces is required.

For the RS II model\cite{RS2}, it has been shown that 
gravity is localized on the brane, which is shown by decomposing
the full graviton propagator\cite{GT,GKR}. 
Here, a localized source induces a localized 
field, which diminishes as one goes toward the AdS 
horizon\cite{GT,GKR,emparan}. And, 
the brane bending effect in the existence of matter on the brane is 
shown to be crucial for consistency of the linearized 
approximation\cite{GKR} and is necessary to reproduce the 4D 
Eintein gravity on the brane\cite{GT,GKR}. Discussions related to the AdS/CFT 
correspondence in the RS II model also have been studied in the
literature\cite{gubser,GKR,GK,duff,nojiri,nunez} from which one can see
that the RS II model is described by the visible matter gravitationally 
coupled to CFT with a cutoff.

It is important to check the consistency of linearized gravity after 
the inclusion of the Gauss-Bonnet term which is called the 
Einstein-Gauss-Bonnet(EGB) theory.
The Gauss-Bonnet term is a topological term in $D=4$\cite{zwie,zumino} 
and it does not affect the graviton propagator even for the
$D>4$ flat spacetime background\cite{zwie}.  
Since there are no higher order derivatives induced from variations of the 
Gauss-Bonnet term other than the second, it seems that there 
is no ghost problem in the low energy limit of the EGB theory. 
Therefore, it is reasonable to take the next leading curvature terms in the 
action as the Gauss-Bonnet term. For the 
RS II model in the EGB theory, it is shown that 
localization of gravity also arises around the RS type static 
background in view of the graviton mass spectrum\cite{KKL}. 
The difference from the original RS II model is that it seems that 
the localized gravity appears on the negative tension brane for the 
static backgrounds allowed\cite{KKL}. In this paper, 
considering the full graviton propagator and the brane bending effect 
in the existence of the Gauss-Bonnet term, we show that excitation of 
ghost particles on the brane is a universal phenomenon in the low energy 
limit, regardless of our static solution backgrounds. This result is 
different from that in \cite{deser}, where it was shown that only one 
of static solutions could excite ghosts in the EGB theory.
Several extensions of the EGB theory with a 5D 
dilaton field  and the general higher order curvature terms 
were considered for discrete or smooth domain wall solutions in the 
subsequent papers\cite{nojiri,extend1,kaku1,extend2}.

In this paper, we obtain the full 5D graviton propagator in the RS II 
model with the Gauss-Bonnet term and a manifest expression for 
localization of gravity which can be shown by 
the decomposition of the graviton propagator. Because the 
gauge choice of the metric perturbation is non-trivial  
with matter on the brane, the simplest gauge condition (the so-called 
Randall-Sundrum gauge condition\cite{RS2}) considered 
without matter is not valid any more. Nonetheless, in order to maintain 
the RS gauge 
even with matter on the brane, it has been shown that the brane is seen to be 
bent along the extra dimension\cite{GT,CGR,GKR,csaki}. 
Without the RS condition the
gauge degrees of freedom are not completely removed, and
in this case the introduction of the brane bending 
mode\cite{GT,CGR,GKR,csaki} becomes crucial to choose the RS gauge and cancel 
the unphysical scalar degree of freedom of the tensor structure of 5D massless 
graviton.
However, elimination of the scalar degree of freedom 
is incomplete in the existence of the Gauss-Bonnet term, and thus the 
4D Einstein gravity should be modified by the residual scalar degree. 
In that case, the bending of light travelling near the Sun results in 
the different value as the factor $\bigg(1-\frac{2}{3}\beta\bigg)^{-1}$ 
of that in the 4D Einstein gravity. (Here we notice that 
$\beta$ is the dimensionless 
quantity given in terms of the Gauss-Bonnet coefficient and bulk 
cosmological constant as we will see later.) And we also investigate 
the stability of the static backgrounds from the amplitude of one massless 
graviton exchange and the Newtonian potential on the brane. 
As a result, the additional RS type 
solution is always unstable under perturbations because it excites 
ghosts giving rise to repulsive gravitational potentials between 
positive matter sources. On the other hand, for the static solution 
connected with the original RS solution, if the source relation is given as 
$S_{11}+S_{22}\neq 0$, the ghost-free condition requires that
the sign of the Gauss-Bonnet coefficient($\alpha$) should be negative 
in our convention, which is consistent with string amplitude 
computations\cite{tsey}, but otherwise it also allows the positive 
sign of $\alpha$.
And the condition of obtaining the Newtonian gravity is necessary for the 
ghost-free condition if $S_{11}+S_{22}\neq 0$, but otherwise just sufficient.  

In Sec. II, we present equations of motion of the EGB theory in the RS
background. In Secs. III and IV we obtain Green's function and gravitational
potential, respectively. In Sec. V solutions of the metric components
are derived, and in Sec. VI we comment on the possibility of the 
light bending effect in the EGB theory. Sec. VII is a conclusion. 

\section{The Einstein-Gauss-Bonnet theory in the Randall-Sundrum background}

The RS metric of 5D warped spacetime consistent with the orbifold 
symmetry $y\rightarrow -y$\cite{RS1,RS2} is given by
\begin{eqnarray}
ds^2&=&e^{-2k|y|}\eta_{\mu\nu}dx^\mu dx^\nu +dy^2 \label{rsmetric} \\
&\equiv&\bar{g}_{MN}dx^M dx^N
\end{eqnarray}
where $y$ is the fifth coordinate with $y\in (-\infty,\infty)$, 
of which just the half $[0,\infty)$ will be considered and $k$ 
is the AdS curvature scale.
Even with the Gauss-Bonnet term added in the 5D Einstein gravity 
(see Appendix A for the model setup), the static solutions are 
shown to have the same form as the RS metric, except the definition 
of $k$\cite{KKL};
\begin{eqnarray}
k=k_{\pm}\equiv \bigg(\frac{M^2}{4\alpha}\bigg[1\pm 
\sqrt{1+\frac{4\alpha\Lambda_b}{3M^5}}\bigg]\bigg)^{1/2} \label{curvature}
\end{eqnarray}
where $M$, $\alpha$ and $\Lambda_b$ are the 5D fundamental scale, 
the dimensionless parameter of the Gauss-Bonnet term and the bulk 
cosmological constant, respectively. 

Let us expand the EGB theory around the RS metric, 
\begin{equation}
g_{MN}=\bar{g}_{MN}+h_{MN},\label{pert} 
\end{equation}
with the RS gauge condition \cite{RS2},
\begin{equation} 
h_{55}=h_{5\mu}=0,\ \  (\rm Gaussian\ normal(GN)\ condition)
\end{equation}
and 
\begin{equation}
h_\mu^\mu=\partial^\mu h_{\mu\nu}=0\ (\rm 4D\ transverse\ 
traceless(TT)\ condition). 
\end{equation}
Then we obtain the following linearized 
equations of motion for the case of a single brane without 
matter\cite{KKL},
\begin{eqnarray}
&\bigg[-\frac{1}{2}\bigg(1-\frac{4\alpha k^2}{M^2}
+\frac{8\alpha k}{M^2}\delta(y)\bigg)
\partial_{\lambda}^2 e^{2k|y|}-\frac{1}{2}\bigg(1-\frac{4\alpha k^2}
{M^2}\bigg)\partial_y^2
+\frac{8\alpha k^2}{M^2}sgn(y)\delta(y)\partial_y\nonumber\\ 
&+2k^2\bigg(1-\frac{4\alpha k^2}{M^2}\bigg)-2k\bigg(1
-\frac{12\alpha k^2}{M^2}\bigg)
\delta(y)\bigg]h_{\mu\nu}(x,y)=0 \label{eqnomatter}.
\end{eqnarray}
The general higher curvature terms tend to give rise to the delocalization 
of gravity\cite{kaku,kaku1}, but investigation of the graviton mass 
spectrum of the above equation shows\cite{KKL} 
that there is one bound state of massless 
graviton on the brane which is identified as the 4D graviton state, 
while the Kaluza-Klein modes 
give rise to small corrections to the Newtonian gravity for the length 
scale larger than the fundamental length scale. The fact that the 
localization of gravity is allowed in the EGB theory
has been shown previously in \cite{kaku,kaku1} by 
integrating the action with respect to the extra dimension. 
In general, however, with or without the Gauss-Bonnet term, the 
metric does not satisfy the RS gauge condition on the brane with 
matter unless the matter energy-momentum tensor satisfies a 
specific condition, $T_\mu\,^\mu =0$, as we will see later. Thus, 
it is necessary to start with the more general gauge other than the 
RS gauge\cite{GKR,MK1,MK2,kaku} or find out the metric junction 
condition on the brane by rewriting the Einstein's equations in 
terms of the extrinsic curvature tensor\cite{GT,DD}. 

There exists an ambiguity of gauge 
choice for the metric perturbation around the RS background. 
For the massless state in the Kaluza-Klein reduction of the RS model, 
the 5D TT condition, $\partial^M h_{MN}=0=h_M\,^M$, leads us to be 
left with 5 degrees of freedom(DOF) of massless state after the 
residual gauge transformations, but with 6 DOFs without the trace 
free condition. The 5(6) DOFs are composed of two massless traceless 
symmetric tensor modes, two massless vector modes and one(two) 
massless scalar mode(s) in the 4D sense. Particularly, for the 
Gaussian normal condition, two vector modes and one scalar mode are 
decoupled while two tensor modes remain for 5D TT condition [Case (1)], 
and two tensor modes and one scalar mode remain for 5D transverse 
condition [Case (2)]. Case (1) does not satisfy the consistent 
boundary condition for the existence of matter on the brane, but 
it may do so if we make loose the GN condition to: $h_{5\mu}=0$ and 
$h_{55}(=- h_\mu^\mu)\neq 0$. On the other hand, 
for Case (2), which can be 
shown to satisfy the boundary condition with matter on the brane, 
there exists an additional unphysical scalar degree of freedom, which 
is shown to be cancelled by the fifth coordinate transformation 
(brane bending) maintaining the metric as a GN form\cite{GT,csaki}. 

When we choose just the Gaussian normal condition without the 4D 
transverse traceless condition, the linearized equations of motion 
for $h_{\mu\nu}$ and $h\equiv h_\mu^\mu$ with matter on the brane 
are given as follows (see Appendix A for the details),
\begin{eqnarray}
&\bigg[-\frac{1}{2}\bigg(1-\frac{4\alpha k^2}{M^2}
+\frac{8\alpha k}{M^2}\delta(y)\bigg)
\partial_{\lambda}^2 e^{2k|y|}-\frac{1}{2}
\bigg(1-\frac{4\alpha k^2}{M^2}\bigg)\partial_y^2
+\frac{8\alpha k^2}{M^2}sgn(y)\delta(y)\partial_y\nonumber\\
&+2k^2\bigg(1-\frac{4\alpha k^2}{M^2}\bigg)
-2k\bigg(1-\frac{12\alpha k^2}{M^2}\bigg)
\delta(y)\bigg](h_{\mu\nu}-\eta_{\mu\nu}h)\label{min}\\
&+\frac{1}{2}\bigg(1-\frac{4\alpha k^2}{M^2}
+\frac{8\alpha k}{M^2}\delta(y)\bigg)e^{2k|y|}
(2\partial_{(\mu}\partial^\lambda h_{\nu)\lambda}
-\partial_\mu \partial_\nu h -\eta_{\mu\nu}
\partial_\lambda\partial_\rho h^{\lambda\rho})=
M^{-3}T_{\mu\nu}, \label{munucomp}\nonumber
\end{eqnarray}
\begin{equation}
\bigg(1-\frac{4\alpha k^2}{M^2}\bigg)
\delta^{(1)}R_{5\mu}=M^{-3}T_{5\mu},\label{ymucomp}\\
\end{equation}
\begin{equation}
\bigg(1-\frac{4\alpha k^2}{M^2}\bigg)\delta^{(1)}
\bigg(R_{55}-\frac{1}{2}g_{55}R\bigg)=M^{-3}T_{55},\label{yycomp}
\end{equation}
where 
\begin{eqnarray}
T_{\mu\nu}&\equiv& S_{\mu\nu}(x)\delta(y)\label{tmumu}, \\
\delta^{(1)}R_{5\mu}&=&\frac{1}{2}\partial_y\bigg[e^{2k|y|}(\partial^\lambda h_{\mu\lambda}-\partial_\mu h)\bigg],\label{tymu} \\
\delta^{(1)}\bigg(R_{55}-\frac{1}{2}g_{55}R\bigg)&=&-\frac{1}{2}e^{4k|y|}\partial^\mu\partial^\nu (h_{\mu\nu}-\eta_{\mu\nu}h)-\frac{3}{2}k\, sgn(y)\partial_y(e^{2k|y|}h)\label{tyy}.
\end{eqnarray}
Here, $\delta^{(1)}$ denotes the linear perturbation.
If we take the trace on both sides of the $(\mu\nu)$-component 
of Eq.~(\ref{min}) and use the $(55)$-component of Eq.~(\ref{yycomp}), 
we obtain the equation of motion for the region $y>0$ and the boundary 
condition at $y=0$ for $h$ as follows,
\begin{eqnarray}
&\partial_y\bigg[e^{-2ky}\partial_y (e^{2ky}h)\bigg]
=\frac{2}{3}M^{-3}\bigg(1-\frac{4\alpha k^2}{M^2}\bigg)^{-1}
\bigg(T_\mu\,^\mu-2 e^{-2ky}T_{55}\bigg), \label{eqh}\\
&e^{-2ky}\partial_y (e^{2ky}h)|_{y=+0}=\frac{1}{3}M^{-3}
\bigg(1-\frac{4\alpha k^2}{M^2}\bigg)^{-1}\bigg[S_\mu\,^\mu(x)-2
\bigg(\frac{8\alpha k}{1-4\alpha k^2/M^2}\bigg)T_{55}(0)\bigg]\label{bch}.
\end{eqnarray}
The $h$ equation can be rewritten by using the conservation of $T$, i.e.
$\bar{\cov}_M T^{M5}=0$ (where $\bar\cov$ denotes the action on the
background),
\begin{equation}
\partial_y\bigg[e^{-2ky}\bigg(\partial_y (e^{2ky}h)
+\frac{2}{3k}M^{-3}\bigg(1-\frac{4\alpha k^2}
{M^2}\bigg)^{-1}T_{55}\bigg)\bigg]=-\frac{2}
{3k}M^{-3}\bigg(1-\frac{4\alpha k^2}{M^2}\bigg)^{-1}
\partial^\mu T_{\mu 5} \label{eqh2}.
\end{equation}

\section{Green's function}

>From Eq.~(\ref{eqh}) or Eq.~(\ref{eqh2}), 
if $T_\mu\,^\mu\neq 0$ or $\partial^\mu T_{\mu 5}\neq 0$, 
the trace $h$ has the exponentially growing component, 
in which case the linear approximation breaks down. 
So, to cancel the growing component, let us consider a 
deformation of coordinates maintaining the metric as a 
Gaussian normal form\cite{GT,CGR,GKR}; 
\begin{eqnarray}
y^\prime&=&y-\xi^5 (x), \nonumber \\ 
x^{\prime\mu}&=&x^\mu-\xi^\mu(x,y) \nonumber\\
&=&x^\mu+
\frac{1}{2k}e^{2ky}\eta^{\mu\lambda}
\partial_\lambda\xi^5(x)-\hat{\xi}^\mu(x) \label{tranf},
\end{eqnarray}
such that
\begin{equation}
ds^2=g_{\mu\nu}(x,y)dx^\mu dx^\nu+dy^2=g^\prime_{\mu\nu}
(x^\prime,y^\prime){dx^\prime}^\mu {dx^\prime}^\nu
+{dy^\prime}^2\label{gn}. 
\end{equation}
Then, the metric perturbation transforms as 
\begin{equation}
h_{\mu\nu}(x,y)\rightarrow h^\prime_{\mu\nu}
(x,y)=h_{\mu\nu}(x,y)-\frac{1}{k}
\partial_\mu\partial_\nu \xi^5 (x)+e^{-2ky}[2\partial_{(\mu}\hat{\xi}_{\nu )}
(x)-2k\xi^5 (x)\eta_{\mu\nu}]\label{trfmetric}.
\end{equation} 
When we integrate both sides of the Eq.~(\ref{eqh2}) with respect 
to $y$ with the initial condition at $y=0$, Eq.~(\ref{bch}), 
we have the exponentially growing component of $h$ eliminated 
by the gauge choice with 
\begin{equation}
\partial_\mu\partial^\mu\xi^5 (x)=\frac{1}{3}
M^{-3}\bigg(1-\frac{4\alpha k^2}{M^2}\bigg)^{-1}
\bigg[\frac{S_\mu\,^\mu(x)}{2}+\frac{1}{k}\bigg(\frac{1-12\alpha k^2/M^2}
{1-4\alpha k^2/M^2}\bigg)T_{55}(0)-\frac{1}{k}
\int_0^{y_m}dy\,\partial^\mu T_{\mu 5}\bigg]\label{aleq}
\end{equation}
where $y_m$ denotes the range of the matter distribution 
along the $y$ coordinate. For the matter localized on the brane, 
we have the solution for $\xi^5$ as
\begin{equation}
\xi^5(x)=\frac{1}{6}M^{-3}\bigg(1-\frac{4\alpha k^2}{M^2}\bigg)^{-1}
\int d^4 x^\prime G_4(x,x^\prime)S_\lambda^\lambda 
(x^\prime)
\label{alsol}
\end{equation}
where $G_4(x,x^\prime)$ is Green's function for the massless scalar 
in the 4D Minkowski space.  Then, in the primed coordinate, integrating  
Eq.~(\ref{eqh2}) with respect to $y$, we obtain  
\begin{equation}
\partial_y(e^{2ky} h^\prime)=-\frac{2}{3k}M^{-3}
\bigg(1-\frac{4\alpha k^2}{M^2}\bigg)^{-1}
\bigg[T_{55}(y)-e^{2ky}\int_y^{y_m}dy\partial^\mu T_{\mu 5}\bigg].\label{eqh3}
\end{equation}  
On the other hand, we obtain the $(5\mu)$-component in the primed 
coordinate from Eq.~(\ref{ymucomp}),
\begin{equation}
\partial_y(e^{2ky}\partial^\lambda h^\prime_{\mu\lambda})
=\partial_y(e^{2ky}\partial_\mu h^\prime)+2M^{-3}\bigg(1
-\frac{4\alpha k^2}{M^2}\bigg)^{-1}T_{5\mu}\label{ymucomp2}.
\end{equation} 
Then, with the definition of the following metric, 
\begin{eqnarray}
\bar{h}_{\mu\nu}\equiv h_{\mu\nu}-\frac{1}{2}\eta_{\mu\nu}h\label{redef},
\end{eqnarray}
from Eq.~(\ref{ymucomp2}) and Eq.~(\ref{eqh3}), we always 
have 
\begin{equation}
\partial_y (e^{2ky}\partial^\lambda \bar{h}^\prime_{\mu\lambda})=0. 
\ {\rm for}\  y>y_m.\label{min1}
\end{equation}
[Note that matter does not exist beyond $y_m$ and
$T_{55}(y>y_m)=0.$] Moreover, we can obtain the de Donder gauge condition 
$\partial^\lambda \bar{h}^\prime_{\mu\lambda}=0$ just outside the matter 
distribution by the remaining gauge invariance $\hat{\xi}_\mu$ and the same 
for $y>y_m$ using Eq.~(\ref{min1}).  
Therefore,  when  there exists the matter source only on the brane 
(i.e, $T_{5\mu}=T_{55}=0$), we can always choose the 4D transverse 
traceless gauge $\partial^\lambda {h^\prime}_{\mu\lambda}=h^\prime=0$ 
both in the bulk and on the brane. 

By the way, the boundary condition of the $(\mu\nu)$-component on the 
brane is given from the Eq.~(\ref{min}) as follows,
\begin{eqnarray}
&\bigg[\bigg(1-\frac{12\alpha k^2}{M^2}\bigg)(\partial_y+2k)
+\frac{4\alpha k}{M^2}e^{2ky}\partial^2_\lambda\bigg]
(h_{\mu\nu}-\eta_{\mu\nu}h)|_{y=+0}\nonumber \\
&-\frac{4\alpha k}{M^2}e^{2ky}(2\partial_{(\mu}
\partial^\lambda h_{\nu)\lambda}-\partial_\mu\partial_\nu 
h-\eta_{\mu\nu}\partial_\lambda\partial_\rho h^{\lambda\rho})|_{y=+0}
=-M^{-3}S_{\mu\nu}\label{bcmunu}.
\end{eqnarray}
Then, with the metric $\bar{h}_{\mu\nu}$,
we can rewrite the above boundary condition as follows,
\begin{eqnarray}
&\bigg(1-\frac{12\alpha k^2}{M^2}\bigg)(\partial_y+2k)
\bigg(\bar{h}_{\mu\nu}+\frac{1}{2}\eta_{\mu\nu}\bar{h}\bigg)|_{y=+0}
\nonumber\\
&+\frac{4\alpha k}{M^2}e^{2ky}(\partial^2_\lambda\bar{h}_{\mu\nu}
-\partial_\mu\partial^\lambda \bar{h}_{\nu\lambda}
-\partial_\nu\partial^\lambda \bar{h}_{\mu\lambda}
+\eta_{\mu\nu}\partial^\lambda\partial^\rho 
\bar{h}_{\lambda\rho})|_{y=+0}=-M^{-3}S_{\mu\nu}\label{bcmunu2}
\end{eqnarray}
which becomes in terms of $\bar h^\prime_{\mu\nu}$,
\begin{eqnarray}
&\bigg(1-\frac{12\alpha k^2}{M^2}\bigg)(\partial_y+2k)
\bigg(\bar{h}^\prime_{\mu\nu}+\frac{1}{2}\eta_{\mu\nu}\bar{h}^\prime\bigg)|_{y=+0}
\nonumber\\
&+\frac{4\alpha k}{M^2}e^{2ky}(\partial^2_\lambda\bar{h}^\prime_{\mu\nu}
-\partial_\mu\partial^\lambda \bar{h}^\prime_{\nu\lambda}
-\partial_\nu\partial^\lambda \bar{h}^\prime_{\mu\lambda}
+\eta_{\mu\nu}\partial^\lambda\partial^\rho 
\bar{h}^\prime_{\lambda\rho})|_{y=+0}\nonumber \\
&=-M^{-3}S_{\mu\nu}-2\bigg(1-\frac{4\alpha k^2}{M^2}\bigg)
(\partial_\mu\partial_\nu-\eta_{\mu\nu}\partial^2_\lambda)\xi^5 
\label{barbc}.
\end{eqnarray}
When $\partial^\mu \bar{h}^\prime_{\mu\nu}=0=h^\prime$ [Note the definitions
given in Eqs.~(\ref{trfmetric}) and (\ref{redef}).] is chosen on the brane 
by the gauge shift of $\hat{\xi}_\mu$,  
we obtain the boundary condition in the primed coordinate,
\begin{eqnarray}
&\bigg(1-\frac{12\alpha k^2}{M^2}\bigg)(\partial_y+2k)
h^\prime_{\mu\nu}|_{y=+0}+\frac{4\alpha k}{M^2}e^{2ky}
\partial^2_\lambda h^\prime_{\mu\nu}|_{y=+0}\nonumber\\
&=-M^{-3}S_{\mu\nu}-2\bigg(1-\frac{4\alpha k^2}{M^2}\bigg)
(\partial_\mu\partial_\nu-\eta_{\mu\nu}\partial^2_\lambda)\xi^5 
\label{bcmunu3}.
\end{eqnarray}
Then, by using the Eq.~(\ref{alsol}), it is shown that the trace of 
the above equation is consistent with the traceless condition $h^\prime=0$. 
Note that the transverse condition $\partial^\mu h^\prime_{\mu\nu}=0$
implies the conservation of the 4D energy-momentum tensor:
$\partial^\mu S_{\mu\nu}=0$.

Consequently, considering the Eq.~(\ref{eqnomatter}) in the RS gauge for the 
case without matter on the brane, the linearized equations of motion for the 
brane with matter becomes in the RS gauge, 
\begin{eqnarray}
&\bigg{[}\bigg(1-\frac{4\alpha k^2}{M^2}+\frac{8\alpha k}
{M^2}\delta(y)\bigg)
\partial_{\lambda}^2 e^{2k|y|}+\bigg(1-\frac{4\alpha k^2}{M^2}\bigg)
\partial_y^2
-\frac{16\alpha k^2}{M^2}sgn(y)\delta(y)\partial_y \nonumber \\
&-4k^2\bigg(1-\frac{4\alpha k^2}{M^2}\bigg)+4k\bigg(1-\frac{12\alpha k^2}{M^2}
\bigg)
\delta(y)\bigg]{h^\prime}_{\mu\nu}(x,y)=-2M^{-3}
\Sigma_{\mu\nu}(x)\delta(y),
\end{eqnarray}\label{munucomp2}
where
\begin{equation}
\Sigma_{\mu\nu}\equiv S_{\mu\nu}+2M^3\bigg(1-\frac{4\alpha k^2}{M^2}\bigg)
(\partial_\mu\partial_\nu-\eta_{\mu\nu}\partial^2_\lambda)\xi^5\label{newem}. 
\end{equation}
Therefore, in the RS gauge for which the brane is located at $y=-\xi^5(x)$, 
we can obtain the metric perturbation in terms of the 
graviton propagator as
\begin{eqnarray}
h^\prime_{\mu\nu}(x,y)&=&-2M^{-3}\int d^4 x^\prime \int_0^\infty dy^\prime\, 
G_5(x,y;x^\prime, y^\prime)\Sigma_{\mu\nu}(x^\prime)\delta(y^\prime) 
\nonumber \\
&=&-M^{-3}\int d^4 x^\prime G_5(x,y;x^\prime,0)\Sigma_{\mu\nu}
(x^\prime) \nonumber \\
&=&-M^{-3}\int d^4 x^\prime G_5(x,y;x^\prime, 0)\bigg[S_{\mu\nu}(x^\prime)
-\frac{1}{3}\eta_{\mu\nu}S_\lambda^\lambda(x^\prime)+\frac{1}{3}
\frac{\partial_\mu\partial_\nu}{\partial^2_\rho}S_\lambda^\lambda(x^\prime)
\bigg]\label{pertprime}, 
\end{eqnarray}
where Eq.~(\ref{alsol}) is used in the last line and 
we note that the graviton propagator is well defined for 
either side of the brane, i.e. valid for the half of the bulk space $y>0$
and the other half by the $Z_2$ symmetry. However, the metric 
$h^\prime_{\mu\nu}$ is not appropriate for observers on the brane since the 
brane becomes bent as $y=-\xi^5(x)$ in the RS gauge. Nonetheless, there exists
a simpler gauge for observers on the brane in which case the brane is located 
at $y=0$ as before the coordinate transformation. It will be shown in Sec. V. 
Then, the equation of motion for the graviton propagator is given by 
\begin{eqnarray}
&\bigg{[}\bigg(1-\frac{4\alpha k^2}{M^2}+\frac{8\alpha k}{M^2}
\delta(y)\bigg)
\partial_{\lambda}^2 e^{2ky}+\bigg(1-\frac{4\alpha k^2}{M^2}\bigg)
\partial_y^2
-\frac{16\alpha k^2}{M^2}sgn(y)\delta(y)\partial_y \nonumber \\
&-4k^2\bigg(1-\frac{4\alpha k^2}{M^2}\bigg)
+4k\bigg(1-\frac{12\alpha k^2}{M^2}
\bigg)
\delta(y)\bigg]G_5(x,y;x^\prime,y^\prime)=\delta^{(4)}(x-x^\prime)
\delta(y-y^\prime)\label{greeneq}.
\end{eqnarray}
Now we can decompose the graviton propagator into Fourier modes as 
\begin{eqnarray}
G_5(x,y;x^\prime, y^\prime)=\int\frac{d^4 p}{(2\pi)^4}
e^{ip(x-x^\prime)}G_p(y,y^\prime),\label{fourier} 
\end{eqnarray}
and the equation of motion for the Fourier 
mode $G_p(y,y^\prime)$ is given by
\begin{eqnarray}
&\bigg{[}-\bigg(1-\frac{4\alpha k^2}{M^2}+\frac{8\alpha k}{M^2}
\delta(y)\bigg)
p^2 e^{2ky}+\bigg(1-\frac{4\alpha k^2}{M^2}\bigg)\partial_y^2
-\frac{16\alpha k^2}{M^2}sgn(y)\delta(y)\partial_y \nonumber \\
&-4k^2\bigg(1-\frac{4\alpha k^2}{M^2}\bigg)+4k\bigg(1
-\frac{12\alpha k^2}{M^2}\bigg)
\delta(y)\bigg]G_p(y,y^\prime)=\delta(y-y^\prime)\label{fgreeneq}.
\end{eqnarray}

\section{Gravitational potential}

By a change of variable $z\equiv \frac{e^{ky}}{k}$, the following equation 
of motion and the boundary condition are satisfied by the propagator,
\begin{eqnarray}
&(z^2\partial^2_z+z\partial_z-p^2 z^2-4)G_p(z,z^\prime)=\bigg(1
-\frac{4\alpha k^2}{M^2}\bigg)^{-1}\frac{z}{k}\delta(z-z^\prime)
\label{greeneq2}  \\
&(z\partial_z+2-\beta p^2 z^2)G_p(z,z^\prime)|_{z=1/k}=0
\label{greenbc}
\end{eqnarray}
where
\begin{eqnarray}
\beta \equiv \frac{4\alpha k^2/M^2}{1-12\alpha k^2/M^2}\label{beta}.
\end{eqnarray}
The solution for the graviton propagator is composed of the Bessel 
functions of order 2, ($J_2(q/k)$ and $Y_2(q/k)$, $q^2=-p^2$) and 
it must satisfy Eq.~(\ref{greenbc}) and the boundary conditions at 
$z=z^\prime$,
\begin{eqnarray}
G_<|_{z=z^\prime}&=&G_>|_{z=z^\prime}\label{conti}, \\
\partial_z(G_>-G_<)|_{z=z^\prime}&=&\bigg(1-\frac{4\alpha k^2}{M^2}\bigg)^{-1}\frac{1}{kz^\prime}\label{deriv}
\end{eqnarray}
where $>(<)$ indicates $z$ larger(lesser) than $z^\prime$.

For $z<z^\prime$, taking into account the Eq.~(\ref{greenbc}), we have 
\begin{eqnarray}
G_<(z,z^\prime)&=&A(z^\prime)\bigg[\bigg(Y_1(q/k)+\frac{\beta q}{k}
Y_2(q/k)\bigg)J_2(qz)-\bigg(J_1(q/k)+\frac{\beta q}{k}J_2(q/k)\bigg)
Y_2(qz)\bigg] \nonumber \\
&=&iA(z^\prime)\bigg[\bigg(J_1(q/k)+\frac{\beta q}{k}
J_2(q/k)\bigg)
H^{(1)}_2(qz)-\bigg(H^{(1)}_1(q/k)+\frac{\beta q}{k}H^{(1)}_2(q/k)
\bigg)J_2(qz)\bigg]\label{sol1}
\end{eqnarray}
where $H^{(1)}_{1,2}=J_{1,2}+iY_{1,2}$ is the first Hankel function.
On the other hand, let us turn to the range of $z>z^\prime$. Requiring 
that positive frequency gravitational waves be ingoing into the AdS 
horizon\cite{GKR}, which implies that there is no radiation reemitted 
from the AdS horizon, we also obtain  
\begin{eqnarray}
G_>(z,z^\prime)=B(z^\prime)H^{(1)}_2(qz)\label{sol2}.
\end{eqnarray}
>From the matching conditions Eq.~(\ref{conti}) and Eq.~(\ref{deriv}), 
we get the graviton propagator in momentum space as follows,
\begin{eqnarray}
G_p(z,z^\prime)&=&\bigg(1-\frac{4\alpha k^2}{M^2}\bigg)^{-1}
\bigg(\frac{i\pi}{2k}\bigg)\bigg[\bigg(\frac{J_1(q/k)
+\frac{\beta q}{k}J_2(q/k)}{H^{(1)}_1(q/k)+\frac{\beta q}{k}H^{(1)}_2(q/k)}
\bigg)H^{(1)}_2(qz_<)H^{(1)}_2(qz_>) \nonumber \\
&-&J_2(qz_<)H^{(1)}_2(qz_>)\bigg].\label{greenmom}
\end{eqnarray} 
Then, for the source on the brane, i.e., $z^\prime=\frac{1}{k}$, 
the graviton propagator in coordinate space is given by
\begin{eqnarray}
G_5(x,z;x^\prime, \frac{1}{k})=\bigg(1-\frac{4\alpha k^2}{M^2}\bigg)^{-1}
\int\frac{d^4 p}{(2\pi)^4} e^{ip(x-x^\prime)}\frac{1}{q}
\bigg[\frac{H^{(1)}_2(qz)}{H^{(1)}_1(q/k)+\frac{\beta q}{k}
H^{(1)}_2(q/k)}\bigg]\label{greengen}.
\end{eqnarray}
Furthermore, using the Bessel recursion relation 
$H^{(1)}_2(q/k)=\frac{2k}{q}H^{(1)}_1(q/k)-H^{(1)}_0(q/k)$, we have 
the graviton propagator on the brane decomposed into two
terms, corresponding to the massless mode and the Kaluza-Klein(KK) 
states, respectively,
\begin{eqnarray}
G_5(x,\frac{1}{k};x^\prime, \frac{1}{k})
\equiv\bigg(1-\frac{4\alpha k^2}{M^2}\bigg)^{-1}\bigg[\bigg(\frac{2k}
{1+2\beta}\bigg) G_4(x,x^\prime)\bigg]+G_{KK}(x,x^\prime)
\label{greensum}, 
\end{eqnarray} 
where
\begin{eqnarray}
G_4(x,x^\prime)&\equiv&\int\frac{d^4 p}{(2\pi)^4} 
e^{ip(x-x^\prime)}\frac{1}{q^2}\label{4dgreen},\\
G_{KK}(x,x^\prime)&\equiv& -\left(1-\frac{4\alpha k^2}{M^2}\right)^{-1}
\bigg(\frac{1}{1+2\beta}\bigg)^2 \int\frac{d^4 p}{(2\pi)^4} e^{ip(x-x^\prime)}
\cdot\nonumber\\
&\cdot&\frac{1}{q}
\bigg[\frac{H^{(1)}_0(q/k)}{H^{(1)}_1(q/k)-\bigg(\frac{\beta q/k}
{1+2\beta}\bigg)H^{(1)}_0(q/k)}\bigg]\label{kkgreen}.
\end{eqnarray}
That is, from the above decomposition of the graviton propagator, 
we confirm the localization of gravity on the brane even in the 
existence of the Gauss-Bonnet term, irrespective of the static 
solution backgrounds. However, since there appears the non-trivial 
factor in front of the 4D Minkowski Green function, we should be 
careful with the ghost-like behavior of the resultant Newtonian 
potential, which will be dealed with later. And, we can see that 
the Gauss-Bonnet term affects both massless graviton and KK graviton 
propagators unlike the case in the flat background spacetime\cite{zwie}.

>From the exact form of graviton propagator we found above, 
let us consider the asymptotics of the graviton propagator for a 
static source on the brane. 
At large distances $r\equiv |x-x^\prime|\gg k^{-1}$ or low 
energies ${p}/{k}\ll 1$, the graviton propagator on the brane has 
the following limiting behavior,
\begin{eqnarray}
V(r)&=&\int dt G_5(x,\frac{1}{k};x^\prime,\frac{1}{k})\nonumber \\
&\simeq&\bigg(1-\frac{4\alpha k^2}{M^2}\bigg)^{-1}\int \frac{d^3 p}
{(2\pi)^3}e^{ip(x-x^\prime)}\bigg[-\bigg(\frac{1}{1+2\beta}\bigg)
\frac{2k}{p^2} 
+\bigg(\frac{1}{1+2\beta}\bigg)^2\frac{1}{k}\,ln(p/2k)\bigg]
\label{branelimit} \\
&=&-\bigg(1-\frac{4\alpha k^2}{M^2}\bigg)^{-1}\frac{k}{2\pi r}
\bigg[\bigg(\frac{1}{1+2\beta}\bigg)+\bigg(\frac{1}{1+2\beta}\bigg)^2
\frac{1}{2(kr)^2}\bigg]\nonumber.
\end{eqnarray} 
We note that the above static limit of the graviton propagator is 
very similar to the result of Ref.~\cite{KKL}. Nonetheless, the nontrivial 
factor $(1+2\beta)^{-1}$ as well as the overall factor 
$(1-\frac{4\alpha k^2}{M^2})^{-1}$ in front of the Newton potential, 
${1}/{r}$, show the crucial point such as the ghost-like 
behavior in the existence of the Gauss-Bonnet term. However, even if 
the equation of the graviton propagator does not allow the mode sum in our 
case because of the non-hermitianity of the differential operator, 
we may regard the graviton propagator as the sum of massless mode and 
massive KK modes just as in the RS case\cite{RS2}. 

On the other hand, for $z\gg k^{-1}$ and/or 
$r\equiv |x-x^\prime|\gg k^{-1}$, another static limit of the graviton 
propagator off the brane is given by
\begin{eqnarray}
V(r,z)&=& \int dt G_5(x,z;x^\prime,\frac{1}{k}) \\
&\simeq&\bigg(1-\frac{4\alpha k^2}{M^2}\bigg)^{-1}\bigg(\frac{1}{1
+2\beta}\bigg)
\frac{\pi i}{2k}\int \frac{d^3 p}{(2\pi)^3}e^{ip(x-x^\prime)}
{H_2}^{(1)}(ipz) \\
&\simeq&-\bigg(1-\frac{4\alpha k^2}{M^2}\bigg)^{-1}\bigg(\frac{1}{1
+2\beta}\bigg)\frac{3}{4\pi k}\frac{1}{z^3}\frac{1+\frac{2r^2}{3z^2}}
{\bigg(1+\frac{r^2}{z^2}\bigg)^{3/2}}\label{bulklimit}. 
\end{eqnarray}
As a result, the metric perturbation falls off as $h\sim \frac{1}{z^3}$ 
(for $z\gg r$), which ensures that the perturbative expansion is valid 
towards the AdS horizon\cite{GKR}. And, a localized gravitational 
source produces a localized field, not strong at the horizon.  

\section{Solutions for the metric components}

To find solutions of the initial metric perturbation $h_{\mu\nu}$, 
by using the 
additional freedom $\hat{\xi}_\mu(x)$ in the inverse transformation of the 
metric in Eq.~(\ref{trfmetric}) to set
\begin{eqnarray}
\hat{\xi}_\mu(x)=\partial_\mu\bigg(\frac{1}{2k}\xi^5(x)-\frac{1}{3}M^{-3}
\int d^4 x^\prime G_5(x,\frac{1}{k};x^\prime,\frac{1}{k})\frac{1}
{\partial^2_\rho}S^\lambda_\lambda(x^\prime)\bigg)\label{betafix}, 
\end{eqnarray}
we can have a simple expression for the metric perturbation on the 
brane at $z=\frac{1}{k}$ whose 4D hypersurface is perpendicular to 
the AdS horizon,
\begin{eqnarray}
h_{\mu\nu}(x)=h^{(m)}_{\mu\nu}(x)+h^{(b)}_{\mu\nu}(x)\label{pertunprime}
\end{eqnarray}
where $^{(m)}$ denotes the matter contribution and $^{(b)}$ denotes
the brane bending effect,
\begin{eqnarray}
h^{(m)}_{\mu\nu}(x)&=&-M^{-3}\int d^4 x^\prime G_5(x,\frac{1}{k};x^\prime,
\frac{1}{k})\bigg(S_{\mu\nu}(x^\prime)-\frac{1}{3}
\eta_{\mu\nu}S^\lambda_\lambda(x^\prime)\bigg)\label{matter}
\\
h^{(b)}_{\mu\nu}(x)&=&2k\eta_{\mu\nu}\xi^5(x)\label{bending}.
\end{eqnarray}
The second term of the metric perturbation $h^{(b)}_{\mu\nu}$ in 
Eq.~(\ref{pertunprime}) is due to the brane bending\cite{GT,CGR,GKR,csaki}, 
which is a scalar mode on the brane coming from the trace part 
of matter energy-momentum tensor in Eq.~(\ref{alsol}). We can rewrite 
Eq.~(\ref{pertunprime}) by using 
Eq.~(\ref{greensum}) and Eq.~(\ref{alsol}) as follows,
\begin{eqnarray}
h_{\mu\nu}(x)&=&-M^{-2}_P\int d^4 x^\prime G_4(x,x^\prime)
\bigg[S_{\mu\nu}(x^\prime)-\bigg(\frac{1}{2}+\frac{1}{3}\beta\bigg)
\eta_{\mu\nu}S^\lambda_\lambda(x^\prime)\bigg] \nonumber \\
&-&M^{-3}\int d^4 x^\prime G_{KK}(x,x^\prime)\bigg[S_{\mu\nu}(x^\prime)
-\frac{1}{3} \eta_{\mu\nu}S^\lambda_\lambda(x^\prime)\bigg]\label{pertfinal}
\end{eqnarray}
where brane bending effect contributes only to the
trace part. The effective 4D Planck scale $M_P$ in the limit of 
vanishing $\beta$ or $\alpha$ could be identified as 
\begin{eqnarray}
M^2_P&\equiv& \frac{M^3}{2k}\bigg(1-\frac{4\alpha k^2}{M^2}\bigg)
(1+2\beta)\equiv \frac{1}{16\pi G_N} \nonumber\\
&=&\frac{M^3}{2k}\bigg(1-\frac{4\alpha k^2}{M^2}\bigg)^2
\bigg(1-\frac{12\alpha k^2}{M^2}\bigg)^{-1}.
\end{eqnarray} From 
the above expression for the metric perturbation, we can see 
that as $\beta\rightarrow 0$ or $\alpha\rightarrow 0$, the brane 
bending effect yields the usual factor $\frac{1}{2}$\cite{GT,csaki} 
of the trace part of the 4D graviton propagator, but otherwise the 
4D Einstein gravity could be modified by the additional polarization 
factor $\frac{1}{3}\beta$. The case for the bending of light will 
be treated in Sec. VI. 

Let us consider the amplitude for one massless graviton exchange in
the presence of the source $S_{\mu\nu}$\cite{MK1,MK2}. 
In momentum space, it is given from the first part 
of the metric perturbation Eq.~(\ref{pertfinal}) as follows,
\begin{eqnarray}
{\cal L}^{massless}&=&\frac{1}{4}h_{\mu\nu}S^{\mu\nu} \nonumber \\
&=&\frac{4\pi G_N}{p^2}\bigg[S_{\mu\nu}S^{\mu\nu}-\bigg(\frac{1}{2}
+\frac{1}{3}\beta\bigg)(S_\lambda^\lambda)^2\bigg].\label{ampl}
\end{eqnarray} 
If we take a Lorentz frame in which $p^\mu$ is given as a null vector such as
\begin{eqnarray}
p^0=p^3, \,\,\, p^1=p^2=0, \label{lorentz}
\end{eqnarray}
we obtain the following source relations, from the transverse condition, 
$\partial^\mu S_{\mu\nu}=0$, 
\begin{eqnarray}
S_{00}=S_{03}=S_{33},\,\, S_{01}=S_{13},\,\, S_{02}=S_{23}.\label{source}
\end{eqnarray}  
Therefore, we obtain the amplitude as
\begin{eqnarray}
{\cal L}^{massless}=\frac{4\pi G_N}{-p^2_0+p^2_3}\bigg[|S_{+2}|^2+|S_{-2}|^2
-\frac{1}{3}\beta(S_{11}+S_{22})^2\bigg]
\end{eqnarray}
where the first two terms are due to the exchange of massless spin-2 graviton, 
$S_{\pm 2}\equiv \frac{1}{2}(S_{11}-S_{22})\pm iS_{12}$, 
and the third term comes from the residual 
massless graviscalar. To obtain a positive definite amplitude, 
the conditions, $G_{N}>0$ and $\beta\leq 0$, should be satisfied 
simultaneously. 
\\
\indent
For the $k_+$ solution, the amplitude is always negative since $G_N<0$ so that  
ghost states could be excited near the background. That is, it means that the 
$k_+$ solution is unstable under perturbations and therefore we have to exclude
 it at the perturbative level. 
\\
\indent 
On the other hand, for the $k_-$ solution, if $S_{11}+S_{22}\neq 0$, there 
should be a bound on the bulk parameters such as 
$-\frac{1}{3}<\beta\leq 0$ (equivalently, $\alpha\leq 0$) 
in order to have only 
the positive norm states. Therefore, the allowed sign of $\alpha$, i.e., 
$\alpha<0$ is compatible with string tree amplitude computations\cite{tsey}. 
If we require that $S_{11}+S_{22}=0$, the bound on 
the bulk parameters becomes $\beta>-\frac{1}{3}$; 
$-(5/9)<\frac{4\alpha\Lambda_b}{3M^5}\leq 0$ for $\alpha>0$ and 
any value for $\alpha<0$. 
\\
\indent
By the way, because the tensor structure of the metric perturbation 
will be modified, the above identification of the effective 4D Planck 
scale should appear a little different. For instance, in case of a 
static point source with mass $m$ on the brane, i.e., for the 
energy-momentum tensor 
$S_{\mu\nu}=m\delta_{\mu 0}\delta_{\nu 0}\delta^{(3)}(x)$,
components of the metric perturbation due to the matter and the 
brane bending mode are given by, in accord with Eq.~(\ref{matter}) 
and Eq.~(\ref{bending}), 
\begin{eqnarray}
h^{(m)}_{00}(x)&=&-\frac{2}{3}M^{-3}mV(r), \\
h^{(m)}_{ij}(x)&=&-\frac{1}{3}M^{-3}mV(r)\delta_{ij}, \\
\xi^5(x)&=&\bigg(1-\frac{4\alpha k^2}{M^2}\bigg)^{-1}\frac{M^{-3}m}{24\pi r}.
\end{eqnarray}
where $V(r)$ is the same as given in Eq.~(\ref{branelimit}) and $i,\,j$ 
run from 1 to 3. Therefore, we obtain the approximate 
metric perturbation in the unprimed coordinate 
for a static point source on the brane as the following,
\begin{eqnarray}
h_{00}(x)&\simeq &\frac{2G_N m}{r}\bigg[1-\frac{2}{3}\beta+\frac{2}{3}
\bigg(\frac{1}{1+2\beta}\bigg)^2\frac{1}{(kr)^2}\bigg]\label{fm00a} \\
&=&\frac{2\bar{G}_N m}{r}\bigg[1+\frac{2}{3}
\bigg(1-\frac{2}{3}\beta\bigg)^{-1}\bigg(\frac{1}{1+2\beta}\bigg)^2
\frac{1}{(kr)^2}\bigg],\label{fm00} \\
h_{ij}(x)&\simeq&\frac{2\bar{G}_N m}{r}\bigg[\bigg(\frac{1+\frac{2}{3}\beta}
{1-\frac{2}{3}\beta}\bigg)+\frac{1}{3}\bigg(1-\frac{2}{3}
\beta\bigg)^{-1}\bigg(\frac{1}{1+2\beta}\bigg)^2\frac{1}{(kr)^2}
\bigg]\delta_{ij}, \label{fmij}
\end{eqnarray}
where by the Newton potential $\Phi_N=-\frac{1}{2}h_{00}$, the Newton 
constant is identified as
\begin{eqnarray}
\bar{G}_N &\equiv& \frac{k}{8\pi M^3}\bigg(1-\frac{4\alpha k^2}{M^2}\bigg)^{-1}
\bigg(\frac{1-\frac{2}{3}\beta}{1+2\beta}\bigg)\nonumber \\
&\equiv&\frac{1}{16\pi \bar{M}^2_P},\label{min3}
\end{eqnarray}
where the correct effective 4D Planck scale $\bar M^2_P$ is given by
\begin{eqnarray}
\bar{M}^2_P &\equiv&\bigg(1-\frac{2}{3}\beta\bigg)^{-1}M^2_P\nonumber \\
&=&\frac{M^3}{2k}\bigg(1-\frac{4\alpha k^2}{M^2}\bigg)^2
\bigg(1-\frac{44\alpha k^2}{3M^2}\bigg)^{-1}.\label{min4} 
\end{eqnarray}
\noindent
In the Eq.~(\ref{fm00a}), the first term of $\frac{1}{r}$ potential stems from 
the spin-2 graviton while the second term does from the residual graviscalar. 
But, since the two contributions are of the same behavior as $\frac{1}{r}$, we 
should identify the Newton constant from the sum of the two as in 
Eq.~(\ref{fm00}).  

For the $k_+$ solution ($k=k_+$ in Eq.~(\ref{min3})), 
the Newton constant $G_N$ would be negative 
and $1-\frac{2}{3}\beta>0$ always, which might give rise to repulsive 
forces of massless and massive gravitons.  
That is, the $k_+$ solution would give rise to the anti-gravity so that it is  
shown not to be allowed under perturbations again.

The instability of the 
$k_+$ solution can be shown from the weak energy condition on
the background matter: $T^{(0)}_{MN}\xi^M\xi^N\geq 0$ 
for any null vector $\xi^M$ \cite{FGPW,MK2}. 
We assume that the background matter 
$T^{(0)}_{MN}$ localized on the brane (i.e., $T^{(0)}_{55}=
T^{(0)}_{5\mu}=0$) produces the metric with 4D Poincar$\acute{\rm e}$ 
invariance as follows,
\begin{eqnarray}
ds^2&=&\bar{g}_{MN}dx^M dx^N \nonumber \\
&=&e^{-2\sigma(y)}\eta_{\mu\nu}dx^\mu dx^\nu+dy^2\ . \label{bgmetric} 
\end{eqnarray}
>From the weak energy condition and the modified 
Einstein's equation, $T^{(0)}_{MN}=M^3(\bar{G}_{MN}+\bar{H}_{MN})
+\Lambda_b \bar{g}_{MN}$, we have the following non-negative condition 
for the background metric (compare with the result given in 
Ref.~\cite{kaku1}),
\begin{eqnarray}
\sigma^{\prime\prime}\bigg(1-\frac{4\alpha (\sigma^\prime)^2}{M^2}\bigg)
\geq 0.\label{wec}
\end{eqnarray}
Therefore, for $\sigma^{\prime\prime}\geq 0$ for the localization of 
gravity as in the RS case, we have the condition $1-\frac{4\alpha 
(\sigma^\prime)^2}{M^2}=\mp \sqrt{1+(4\alpha\Lambda_b/3M^5)}\geq 0$, 
for the $k_\pm$ solutions, respectively. Therefore,
the $k_-$ solution 
is the only 4D flat solution allowed with localized gravity that 
satisfies the weak energy condition. On the other hand, the $k_+$ 
solution violates the weak energy condition for localized gravity.

On the other hand, for the $k_-$ solution,
the attractive normal gravitational potential is allowed for the range of 
$\alpha$ as $-0.47<\frac{4\alpha \Lambda_b}{3M^5}\leq 0$ 
for $\alpha>0$ and any value for $\alpha<0$. Therefore, even the $k_-$ 
solution could give rise to a repulsive force in some bulk parameter space 
with $\alpha>0$ even though the weak energy condition is satisfied. 
Recalling that there is a lower bound in the allowed static solution 
space itself such as $-1\leq\frac{4\alpha \Lambda_b}{3M^5}\leq 0$ for 
$\alpha >0$ for the $k_-$ solution, we see that the Gauss-Bonnet 
interaction should be weaker than the bound given in Ref.~\cite{KKL} 
to avoid the anti-gravity by about the half. However, from the consistency
for no ghost states in case of $S_{11}+S_{22}\neq 0$, the 
sign of $\alpha$ should be negative. And, if $S_{11}+S_{22}=0$, 
the ghost-free condition is just necessary for the attractive gravity.

\section{Bending of light and more on the localization problem}

We can also show that the bending of light passing by the
Sun could be modified with the Gauss-Bonnet term. For a source 
in the $xz$ plane on the brane, the bending of light 
travelling in the $z$ direction is described by a Newton-like force law, 
$\ddot{x}=\frac{1}{2}(h_{00}+h_{zz})_{,x}$. Note that corrections
to $h_{00}$ and $h_{zz}$ are different [viz. Eqs.~(\ref{fm00}) 
and (\ref{fmij})], which modifies the bending of
light. If the metric perturbations 
due to the Sun are approximated by those from a point source, 
the bending of light is $\bigg(1-\frac{2}{3}\beta\bigg)^{-1}$ of 
that predicted from the 4D Einstein gravity. Therefore, since the 
maximum deflection angle of light is given as 
$\delta\theta_{max}=(0.95\pm 0.11)1.75^{\prime\prime}$ from the 
experimental measurements\cite{will}, we can get another bound on 
the Gauss-Bonnet coefficient as $-0.20<\frac{4\alpha \Lambda_b}{3M^5}
<1.2$ for the static background connected to the RS solution.
As a result, we can see that 
the ghost bound is compatible with the experimental bound from the 
bending of light. 

Now let us make some comments on quasi-localized gravity 
theories\cite{GRS} with the Gauss-Bonnet term. In the quasi-localized 
gravity, the linearized 4D Newtonian gravity can be reproduced at the 
intermediate scale up to some correction proportional to the decay 
width of metastable graviton while the 5D Einstein gravity remains 
itself both at the short and large distances\cite{GRS}. But, 
because the 4D Newtonian gravity is reproduced by the massive 5D 
graviton, the unwanted polarizations in the tensor structure of 
the 5D graviton propagator cannot give rise to the correct 4D Einstein 
gravity\cite{dvali} and even if the brane bending effect cancels the 
extra polarization at the intermediate scale\cite{csaki}, it gives rise 
to the scalar anti-gravity at the ultra-large distances\cite{ghost}. 
For our case with the Gauss-Bonnet interaction, we cannot escape such 
a ghost problem\cite{ghost,pilo,MK1,MK2}. That is, the gravitational 
potential corresponding to the brane bending mode 
gives a repulsive force even with the Gauss-Bonnet term as follows,
\begin{eqnarray}
V(r)=\int dt \, \bigg(-\frac{1}{2}h^{(b)}_{00}(x)\bigg)=
\frac{kM^{-3}}{24\pi r}\bigg(1-\frac{4\alpha k^2}{M^2}\bigg)^{-1}
\label{scalarpot}. 
\end{eqnarray}
Even though the scalar potential could be attractive for the 
$k_+$ solution, there should exist the more severe ghost problem 
in the 4D graviton sector as already discussed in the 
previous section and moreover the 
weak energy condition of Eq.~(\ref{wec}) would be violated. 
So we argue that there still appears the scalar anti-gravity 
in quasi-localized gravity even with the Gauss-Bonnet interaction. 

\section{Conclusion}

In conclusion, we obtained the full 5D graviton propagator in 
the Einstein-Gauss-Bonnet theory with the Randall-Sundrum
background. From 
the decomposition of the graviton propagator on the brane, we confirmed 
the localized gravity in the Einstein-Gauss-Bonnet theory.
The brane bending effect, even if it is crucial to reproduce the 
right 4D Einstein gravity in the RS model, does not cancel the extra 
polarization of the tensor structure of 5D massless graviton completely 
for the case with the Gauss-Bonnet term. As a result, we can have the bound 
on the Gauss-Bonnet coefficient from relativistic effects such as the 
bending of light. And furthermore, if $S_{11}+S_{22}\neq 0$, the ghost-free 
condition from the amplitude of one graviton exchange requires $\alpha\leq 0$,
which is compatible with string amplitude computations. 
If $S_{11}+S_{22}\neq 0$, the ghost-free condition is sufficient
for giving the correct Newtonian gravity. 

\acknowledgments
This work is supported in part by the BK21 program of Ministry 
of Education, Korea Research Foundation Grant No. KRF-2000-015-DP0072, 
CTP Research Fund of Seoul National University,
and by the Center for High Energy Physics(CHEP),
Kyungpook National University.

\newpage
\appendix

\section{Metric expansions with the Gauss-Bonnet term}

When the Gauss-Bonnet curvature squared interaction is added as 
the additional effective interaction to the 
Einstein-Hilbert term in the RS II model, the 5D action is 
given by the following,
\begin{eqnarray}
S=S_0+S_m
\end{eqnarray}
\begin{eqnarray}
S_0&=&\int d^5x\sqrt{-g}\left( {M^3\over 2}R-\Lambda_b 
+\frac{1}{2}\alpha M R^2
-2\alpha M R_{MN}R^{MN}+\frac{1}{2}\alpha MR_{MNPQ}R^{MNPQ}\right) 
\nonumber \\
&+&\int d^4xdy \delta(y) \sqrt{-g^{(4)}}(-\Lambda),
\label{action} \\
S_m&=&\int d^5 x\sqrt{-g}{\cal L}_m\label{maction}
\end{eqnarray}  
where $g, g^{(4)}$ are the determinants of the metrics in the
bulk and the brane, $M$ is the five dimensional gravitational 
constant, $\Lambda_b$
and $\Lambda$ are the bulk and brane cosmological constants,
$\alpha$ is the effective coupling, and ${\cal L}_m$ is the matter 
Lagrangian on the brane or in the bulk. 

Equations of motion in this EGB theory are,
\begin{eqnarray}
G_{MN}+H_{MN}=M^{-3}(-\Lambda_b g_{MN}+T_{MN})\label{tensoreq}
\end{eqnarray}
where 
\begin{eqnarray}
G_{MN}&\equiv&R_{MN}-\frac{1}{2}g_{MN}R,\label{einstein} \\
H_{MN}&\equiv&\frac{\alpha}{M^2}\bigg[-\frac{1}{2}g_{MN}(R^2-4R^2_{PQ}
+R_{PQST}R^{PQST}) \nonumber \\
&+&2RR_{MN}-4R_{MP}R_N\,^P-4R^K\,_{MPN}R_K\,^P+2R_{MQSP}R_N\,^{QSP}\bigg], 
\label{hcurv}\\
T_{MN}&\equiv&T^{(0)}_{MN}+T^{(m)}_{MN}, \label{emtensor}\\
{T^{(0)}}_M\,^N &\equiv&\frac{\sqrt{-g^{(4)}}}{\sqrt{-g}}\delta(y)
\,diag(-\Lambda,-\Lambda,-\Lambda,-\Lambda,0), \label{bgem}\\
T^{(m)}_{MN}&\equiv& -\frac{2}{\sqrt{-g}}\frac{\delta S_m}{\delta g^{MN}}
\label{mem}.
\end{eqnarray}
For the background metric Eq.~(\ref{bgmetric}),
the nonvanishing components of the Riemann tensor and related things 
are given by
\begin{eqnarray}
\bar{R}_{\mu\nu\lambda\rho}&=&-(\sigma^\prime)^2(\bar{g}_{\mu\lambda}
\bar{g}_{\nu\rho}-\bar{g}_{\mu\rho}\bar{g}_{\nu\lambda}),\label{rmunu} \\ 
\bar{R}_{5\mu 5 \nu}&=&(\sigma^{\prime\prime}-(\sigma^\prime)^2)
\bar{g}_{\mu\nu},\label{r5mu5nu} \\
\bar{R}_{\mu\nu}&=&(\sigma^{\prime\prime}-4(\sigma^\prime)^2)
\bar{g}_{\mu\nu},\label{rmunu}\\
\bar{R}_{55}&=&4(\sigma^{\prime\prime}-(\sigma^\prime)^2),\label{r55}\\
\bar{R}&=&8\sigma^{\prime\prime}-20(\sigma^\prime)^2\label{ricci}.
\end{eqnarray}
Then, considering the metric fluctuation around the RS background
Eq.~(\ref{bgmetric}) such as $g_{MN}=\bar{g}_{MN}+h_{MN}$,
from the general formulae for linearizing the curvature tensors,
\begin{eqnarray}
\delta^{(1)}R_{MN}&=&-\frac{1}{2}\bar{\cov}^2h_{MN}-\bar{R}_{MPNQ}h^{PQ}
+\bar{R}_{(M}\,^P h_{N)P} 
+\bar{\cov}_{(M}\bar{\cov}^P h_{N)P}-\frac{1}{2}\bar{\cov}_M\bar{\cov}_N 
h_P\,^P\label{pertmn}\\
\delta^{(1)}R&=&-\bar{R}_{MN}h^{MN}-\bar{\cov}^2 h_P\,^P+\bar{\cov}_M
\bar{\cov}_N h^{MN}\label{pertr} \\
\delta^{(1)}R^K\,_{MPN}&=&\frac{1}{2}(\bar{R}^K\,_{QPN}h_M\,^Q-\bar{R}^Q
\,_{MPN}h_Q\,^K)
+\bar{\cov}_{[P}\bar{\cov}_{|M|} h_{N]}\,^K-\bar{\cov}_{[P}\bar{\cov}^K 
h_{N]M},\label{pertrie}
\end{eqnarray}
we get the following linear expansions with the Gaussian normal condition, 
$h_{5\mu}=h_{55}=0$; 
\begin{eqnarray}
\delta^{(1)}R_{\mu\nu}&=&-\frac{1}{2}\bar{g}^{PQ}\partial_P\partial_Q 
h_{\mu\nu}-2(\sigma^\prime)^2 h_{\mu\nu} \nonumber \\
&+&e^{2\sigma}\bigg(\partial_{(\mu}\partial^\lambda h_{\nu)\lambda}
-\frac{1}{2}\partial_\mu\partial_\nu h\bigg)+\frac{1}{2}\eta_{\mu\nu}
\sigma^\prime\partial_5 h+
(\sigma^\prime)^2\eta_{\mu\nu}h \label{prmunu}\\
\delta^{(1)}R_{5\mu}&=&\frac{1}{2}e^{2\sigma}\partial_5 
(\partial^\lambda h_{\mu\lambda}-\partial_\mu h)+e^{2\sigma}\sigma^\prime 
(\partial^\lambda h_{\mu\lambda}-\partial_\mu h)=\frac{1}{2}\partial_5 
\bigg[e^{2\sigma}(\partial^\lambda h_{\mu\lambda}-\partial_\mu h)\bigg] 
\label{pr5mu}\\
\delta^{(1)}R_{55}&=&e^{2\sigma}\bigg(-\frac{1}{2}\partial^2_5 h-
\sigma^\prime\partial_5 h-\sigma^{\prime\prime} h\bigg)\label{pr55}\\
\delta^{(1)}R&=&e^{2\sigma}\bigg(e^{2\sigma}\partial_\mu\partial_\nu 
h^{\mu\nu}-\bar{g}^{PQ}\partial_P\partial_Q h+\sigma^\prime\partial_5 
h+(-2\sigma^{\prime\prime}+6(\sigma^\prime)^2)h\bigg)\label{pr},
\end{eqnarray}
and
\begin{eqnarray}
\delta^{(1)}R^\lambda\,_{\mu\rho\nu}&=&\frac{1}{2}(\bar{R}^\lambda
\,_{Q\rho\nu}h_\mu\,^Q-\bar{R}^Q\,_{\mu\rho\nu}h_Q\,^\lambda 
+\bar{\cov}_\rho\bar{\cov}_\mu h_\nu\,^\lambda-\bar{\cov}_\nu\bar{\cov}_\mu 
h_\rho\,^\lambda-\bar{\cov}_\rho\bar{\cov}^\lambda h_{\nu\mu}
+\bar{\cov}_\nu\bar{\cov}^\lambda h_{\rho\mu})\nonumber \\
&=&\frac{1}{2}\bigg[e^{2\sigma}(\partial_\rho\partial_\mu h_\nu
\,^\lambda-\partial_\mu\partial_\nu h_\rho\,^\lambda)
+\bar{g}^{\lambda\sigma}(-\partial_\rho\partial_\sigma h_{\mu\nu}
+\partial_\nu\partial_\sigma h_{\rho\mu}) \nonumber \\
&+&\sigma^\prime (-\eta_{\rho\mu}\partial_5 h_\nu\,^\lambda
+\eta_{\mu\nu}\partial_5 h_\rho\,^\lambda+\bar{g}^\lambda
\,_\rho\partial_5 h_{\mu\nu}-\bar{g}^\lambda\,_\nu\partial_5 
h_{\rho\mu})+2(\sigma^\prime)^2(-\eta_{\rho\mu}h_\nu\,^\lambda
+\eta_{\mu\nu}h_\rho\,^\lambda)\bigg]\label{rie1}\\ 
\delta^{(1)}R^5\,_{\mu 5\nu}&=&\frac{1}{2}(\bar{R}^5\,_{Q 5\nu}h_\mu
\,^Q-\bar{R}^Q\,_{\mu 5\nu} h_Q\,^5+\bar{\cov}_5\bar{\cov}_\mu h_\nu
\,^5-\bar{\cov}_\nu\bar{\cov}_\mu h_5\,^5-\bar{\cov}^2_5 h_{\mu\nu}
+\bar{\cov}_\nu\bar{\cov}^5 h_{5\mu})\nonumber \\
&=&-\frac{1}{2}\partial^2_5 h_{\mu\nu}-\sigma^\prime\partial_5 
h_{\mu\nu}-(\sigma^\prime)^2 h_{\mu\nu}\label{rie2}\\
\delta^{(1)}R^\lambda\,_{5\rho 5}&=&\frac{1}{2}(\bar{R}^\lambda
\,_{Q\rho 5}h_5\,^Q-\bar{R}^Q\,_{5 \rho 5}h_Q\,^\lambda+\bar{\cov}_\rho
\bar{\cov}_5 h_5\,^\lambda-\bar{\cov}^2_5h_\rho\,^\lambda-\bar{\cov}_\rho
\bar{\cov}^\lambda h_{55}+\bar{\cov}_5\bar{\cov}^\lambda h_{\rho 5})
\nonumber \\
&=&e^{2\sigma}\bigg(-\frac{1}{2}\partial^2_5 h_\rho\,^\lambda-\sigma^\prime 
\partial_5 h_\rho\,^\lambda-\sigma^{\prime\prime}h_\rho\,^\lambda\bigg)
\label{rie3}\\
\delta^{(1)}R^\lambda\,_{5\rho\nu}&=&\frac{1}{2}\partial_5 
\bigg[e^{2\sigma}(\partial_\rho h_\nu\,^\lambda-\partial_\nu 
h_\rho\,^\lambda)\bigg]\label{rie4}\\
\delta^{(1)}R^5\,_{55\nu}&=&\delta^{(1)}R^5\,_{555}=0\label{rie5}.
\end{eqnarray}
The general formulae for the higher curvature 
terms are,
\begin{eqnarray}
\delta^{(1)}(R^2)&=&2\bar{R}\,\delta^{(1)}R \label{h1}\\
\delta^{(1)}(R_{MN}R^{MN})&=&2\bar{R}^{PQ}\delta^{(1)}R_{PQ}-2h^{PM}
\bar{R}_{PQ}\bar{R}_M\,^Q \label{h2}\\
\delta^{(1)}(R_{MNPQ}R^{MNPQ})&=&\bar{g}^{MT}\delta^{(1)}
(R_{MNPQ}R_T\,^{NPQ})-h^{MT}\bar{R}_{MNPQ}\bar{R}_T\,^{NPQ}\label{h3}\\
\delta^{(1)}(RR_{MN})&=&\bar{R}_{MN}\,\delta^{(1)}R+\bar{R}
\,\delta^{(1)}R_{MN}\label{h4}\\
\delta^{(1)}(R_{MP}R_N\,^P)&=&\bar{R}_M\,^P \delta^{(1)}R_{NP}
+\bar{R}_N\,^P\delta^{(1)}R_{MP}-h^{PQ}\bar{R}_{MP}\bar{R}_{NQ} 
\label{h5}\\
\delta^{(1)}(R^K\,_{MPN}R_K\,^P)&=&\bar{R}_K\,^P\,\delta^{(1)}R^K\,_{MPN}
+\bar{R}^K\,_M\,^P\,_N\,\delta^{(1)}R_{KP}-h^{PQ}\bar{R}^K\,_{MPN}
\bar{R}_{KQ} \label{h6}\\
\delta^{(1)}(R_{MQSP}R_N\,^{QSP})&=&\bar{R}_S\,^P\,_M\,^Q
\,\delta^{(1)}R^S\,_{PNQ}
+\bar{R}_S\,^P\,_N\,^Q\,\delta^{(1)}R^S\,_{PMQ} 
-h^{QT}\bar{R}_{SPMQ}\bar{R}^{SP}\,_{NT}\label{h7}. 
\end{eqnarray}
Then, imposing the Gaussian normal condition only, 
$h_{55}=h_{5\mu}=0$, linear expansions of the Einstein 
tensor $G_{MN}$, the higher curvature tensor $H_{MN}$ 
and the energy-momentum tensor $T_{MN}$ are given by, 
\begin{eqnarray}
\delta^{(1)}G_{\mu\nu}&=&-\frac{1}{2}\bar{g}^{PQ}
\partial_P\partial_Q (h_{\mu\nu}-\eta_{\mu\nu}h)
+(-4\sigma^{\prime\prime}+8(\sigma^\prime)^2)h_{\mu\nu}
+(\sigma^{\prime\prime}-2(\sigma^\prime)^2)
\eta_{\mu\nu} h  \nonumber \\
&+&\frac{1}{2}e^{2\sigma}(2\partial_{(\mu}\partial^\lambda 
h_{\nu)\lambda}-\partial_\mu\partial_\nu h-\eta_{\mu\nu}
\partial_\lambda\partial_\rho h^{\lambda\rho})\label{gmunu}, \\
\delta^{(1)}G_{5\mu}&=&\frac{1}{2}\partial_y \bigg[e^{2\sigma}
(\partial^\lambda h_{\mu\lambda}-\partial_\mu h)\bigg],
\label{g5mu} \\
\delta^{(1)}G_{55}&=&-\frac{1}{2}e^{4\sigma}
\partial^\mu\partial^\nu (h_{\mu\nu}-\eta_{\mu\nu}h)
-\frac{3}{2}\sigma^\prime\partial_y(e^{2\sigma}h),\label{g55}  
\end{eqnarray}
and
\begin{eqnarray}
\delta^{(1)}H_{\mu\nu}&=&\frac{\alpha}{M^2}\bigg[
\bigg((-2\sigma^{\prime\prime}+2(\sigma^\prime)^2)
\bar{g}^{\lambda\rho}\partial_\lambda\partial_\rho
+2(\sigma^\prime)^2\partial^2_y+4\sigma^{\prime\prime}\sigma^\prime
\partial_y\bigg)(h_{\mu\nu}-\eta_{\mu\nu}h) \nonumber \\
&+&(24\sigma^{\prime\prime}(\sigma^\prime)^2-20(\sigma^\prime)^4)
h_{\mu\nu}+(-12\sigma^{\prime\prime}(\sigma^\prime)^2
+8(\sigma^\prime)^4)\eta_{\mu\nu}h \nonumber \\ 
&+&(2\sigma^{\prime\prime}-2(\sigma^\prime)^2)e^{2\sigma}
(2\partial_{(\mu}\partial^\lambda h_{\nu)\lambda}-\partial_\mu
\partial_\nu h-\eta_{\mu\nu}\partial_\lambda\partial_\rho 
h^{\lambda\rho})\bigg], \label{hmunu}\\
\delta^{(1)}H_{5\mu}&=&-\frac{4\alpha (\sigma^\prime)^2}
{M^2}\delta^{(1)}G_{5\mu}
, \label{hmu5}\\
\delta^{(1)}H_{55}&=&-\frac{4\alpha (\sigma^\prime)^2}{M^2}
\delta^{(1)}G_{55},\label{h55}
\end{eqnarray}
and 
\begin{eqnarray}
\delta^{(1)}T_{\mu\nu}&=&(-\Lambda_b-\Lambda \delta(y))
h_{\mu\nu}+T^{(m)}_{\mu\nu} \nonumber \\
&=&M^3\bigg[6(\sigma^\prime)^2\bigg(1-\frac{2\alpha (\sigma^\prime)^2}
{M^2}\bigg)-3\sigma^{\prime\prime}\bigg(1-\frac{4\alpha (\sigma^\prime)^2}
{M^2}\bigg)\bigg]h_{\mu\nu}+T^{(m)}_{\mu\nu},\label{tmunu} \\
\delta^{(1)}T_{5\mu}&=&T^{(m)}_{5\mu}, \label{t5mu}\\ 
\delta^{(1)}T_{55}&=&T^{(m)}_{55}\label{t55}, 
\end{eqnarray}
where we expressed the bulk and brane cosmological constants 
in terms of the background metric function $\sigma$. 
Consequently, from the modified Einstein's equations, 
$G_{MN}+H_{MN}=M^{-3}T_{MN}$, the linearized equations of motion 
become,
$$
\bigg[-\frac{1}{2}\bigg(1-\frac{4\alpha (\sigma^\prime)^2}{M^2}
+\frac{4\alpha \sigma^{\prime\prime}}{M^2}\bigg)
\partial_{\lambda}^2 e^{2\sigma}-\frac{1}{2}\bigg(1-\frac{4\alpha 
(\sigma^\prime)^2}{M^2}\bigg)\partial_y^2
+\frac{4\alpha \sigma^{\prime\prime}\sigma^\prime}{M^2}\partial_y
$$
$$ 
+2(\sigma^\prime)^2\bigg(1-\frac{4\alpha (\sigma^\prime)^2}{M^2}\bigg)
-\sigma^{\prime\prime}\bigg(1-\frac{12\alpha (\sigma^\prime)^2}{M^2}\bigg)
\bigg](h_{\mu\nu}-\eta_{\mu\nu}h)
$$
\begin{eqnarray} 
+\frac{1}{2}\bigg(1-\frac{4\alpha (\sigma^\prime)^2}{M^2}
+\frac{4\alpha \sigma^{\prime\prime}}{M^2}\bigg)e^{2\sigma}
(2\partial_{(\mu}\partial^\lambda h_{\nu)\lambda}-\partial_\mu 
\partial_\nu h -\eta_{\mu\nu}\partial_\lambda\partial_\rho 
h^{\lambda\rho})=M^{-3}T^{(m)}_{\mu\nu}, \label{amunucomp}
\end{eqnarray}
\begin{eqnarray}
\bigg(1-\frac{4\alpha (\sigma^\prime)^2}{M^2}\bigg)
\delta^{(1)}G_{5\mu}&=&M^{-3}T^{(m)}_{5\mu},\label{aymucomp}\\
\bigg(1-\frac{4\alpha (\sigma^\prime)^2}{M^2}\bigg)\delta^{(1)}G_{55}
&=&M^{-3}T^{(m)}_{55}.\label{ayycomp}
\end{eqnarray}
Apart from terms including $\sigma^{\prime\prime}$, which corresponds 
to the delta function source, the linearized Einstein's equation 
remains intact up to the overall factor $\bigg(1
-\frac{4\alpha (\sigma^\prime)^2}{M^2}\bigg)$ on its left hand side 
even in the existence of the Gauss-Bonnet term. But the overall 
factor is not a positive definite quantity, which implies that a test 
particle might feel a repulsive force (or exchange of ghost particles 
of negative energy) from arbitrary matter source.

\end{document}